\documentclass[a4paper,10pt,twoside]{cpc-hepnp}

\usepackage{multicol}
\usepackage{graphicx}
\usepackage{booktabs}
\usepackage{amssymb,bm,mathrsfs,bbm,amscd}
\usepackage[tbtags]{amsmath}
\usepackage{lastpage}
\usepackage{CJK}

\begin{document}



\title{Kinematic Explanation of Massless Particles Only Having Two Helicity States
}

\author{%
LIU ChangLi $^{1)}$ \email{liucl78@pku.org.cn}%
\quad GE FengJun $^{2)}$ \email{ge\_fengjun@iapcm.ac.cn (Corresponding author is Ge.)} %
}
\maketitle

\address{%
Institute of Applied Physics and Computational Mathematics, Beijing China, 100094
}

\begin{abstract}
Why massless particles, for example photons, can only have two helicity one-particle states is the main subject of this work. As we know, the little group which describes massive particle one-particle states' transformations under the Lorentz transformation is SO(3), while the little group describing massless states is ISO(2). In this paper, a method which is different from In\"{o}n\"{u}-Wigner contraction is used to contract SO(3) group to ISO(2) group. We use this contraction method to prove that the particle can only have two helicity one-particle states from the perspective of kinematics, when the particle mass tends to zero. Our proof is different from the dynamic explanation in the existing theories.
\end{abstract}




\begin{multicols}{2}

\section{Introduction}

In Wigner's paper \cite{wigner39} and Weinberg's \emph{The Quantum Theory of Fields} \cite{weinberg}, the one-particle states are defined as the common eigenvectors of four-momentum operator and spin operator. The transformation of massive one-particle states under the Lorentz transformation is as follows (Eq.2.5.23 in Ref.\cite{weinberg}):
\begin{equation} \label{eq:massive1}
U\left( \Lambda  \right){\Psi _{p,\sigma }} = \sqrt {\frac{{{{\left( {\Lambda p} \right)}^0}}}{{{p^0}}}} \sum\limits_{\sigma '} {{\Psi _{\Lambda p,\sigma '}}D_{\sigma '\sigma }^j\left( {W\left( {\Lambda ,p} \right)} \right)} \end{equation}
Where, $\Lambda$ is Lorentz coordinate transformation; $U\left( \Lambda  \right)$ is the corresponding unitary transformation of the one-particle states in the Hilbert space; ${\Psi _{p,\sigma }}$ is one-particle states; $p$ is momentum; $\sigma$ is 3-spin quantum number(helicity); and $D_{\sigma '\sigma }^j\left( W \right)$ is the representation of the little group $W$ which keeps the standard momentums ( $k = \left( {0,0,0,E} \right)$, where $E$ is the particle's energy) of the massive particles invariant; $j$ is the spin quantum number, and it can take integer and half-integer. Here, we only discuss the situation of $j>0$. It can be seen that the massive particle totally have $2j+1$ helicity states.

Correspondingly, the transformation of the massless one-particle states is as follows (Eq.2.5.42 in Ref.\cite{weinberg}):
\begin{equation} \label{eq:massless1}
U\left( \Lambda  \right){\Psi _{p,\sigma }} = \sqrt {\frac{{{{\left( {\Lambda p} \right)}^0}}}{{{p^0}}}} {\Psi _{\Lambda p,\sigma }}\exp \left( {i\sigma \theta \left( {\Lambda ,p} \right)} \right)
\end{equation}
Where, the representation of the little group which keeps the standard momentums ($k = \left( {0,0,E,E} \right)$) of the massless particles invariant has been written out obviously. The helicity of the massless particles described in Eq.(2) can only have one value (p72,p78 in Ref.\cite{weinberg}). If the system is invariant under parity transformation, the other value can be obtained through space inversion, and thus there are totally two helicity states for massless particles.

The little group representation in Eq.(1) is SO(3) representation, and the little group representation in Eq.(2) is the degenerated ISO(2) representation (in fact, it is the representation of SO(2) group). In Ref.\cite{weinberg}, different methods are used to obtain these two little groups and their representations. In order to obviously reveal the internal physical relationship between massive and massless one-particle states, in this paper we used a method to simultaneously obtain the little groups of massive and massless states and their representations. This method is associated with the mass-energy relation of special relativity, and it is different from In\"{o}n\"{u}-Wigner contraction\cite{gilmore}. The method may be proposed by Marek Czachor\cite{czachor} or YS Kim\cite{kim}.

In the existing theoretical system, generally the researchers explain why the massless particles (photons) can only have two helicity states from the dynamic perspective \cite{greiner}. Namely, it describes that the scalar photons and longitudinal photons are nonphysical, and these two kinds of photons cancel in the average value of any physical quantity and only leave the action of the two transverse photons whose helicity are $\pm1$. In the latter part of the paper, we use the contraction method to prove the experimental fact that the helicity of the massless particles has only two values from the perspective of kinematics. The proof proposed by us is different from the dynamic method in existing theories\cite{greiner}.

The Lorentz metric used in this paper is as follows: $g=diag(1,1,1,-1)$, which is  the same as Ref.\cite{weinberg}. Our work is mainly based on the section 2.5 in Ref.\cite{weinberg}.

\section{Method of contracting SO(3) group to ISO(2) group}
The method had been briefly described by Marek Czachor\cite{czachor}. Now, we rewrite it in another way in detail, and it has been generalized to higher-order $SO(n)$ group contraction in the appedix. We know that the selection of the standard momentums is considerable arbitrary. In Ref.\cite{weinberg}, different standard momentums are selected for massive and massless particles. Selecting the same standard momentums for both is a very natural consideration. For massive and massless particles, their standard momentums are both set as $k = \left( {0,0,p,E} \right)$ in our work. For some specific positive energy, from the mass-energy equation  ${E^2} = {p^2} + {m^2}$, it can know that the particles with different masses appear as different 3-component $p$. When $p=0$, $k$ returns to the standard momentum ($\left( {0,0,0,E} \right)$,p66 in Ref.\cite{weinberg}) of the massive one-particle states; when $p=E$, which means that the particle's mass ($m$) is 0, $k$ is the standard momentum ($\left( {0,0,E,E} \right)$,p66 in Ref.\cite{weinberg}) of the massless one-particle states. The standard momentums set in this way can simultaneously discuss massive and massless one-particle states.

In the following, we will solve the expression of the little group $W$ which satisfies  $W_{\ \nu }^\mu {k^\nu } = {k^\mu }$ (that is to say, the little group $W$ keeps the standard momentum  ${k^\mu } = \left( {0,0,p,E} \right)$ invariant). First, $W$ can be split into the product of two Lorentz transformations, namely  $W\left( {\alpha ,\beta ,\theta } \right) = S\left( {\alpha ,\beta } \right)R\left( \theta  \right)$ (p69-70 in Ref.\cite{weinberg}), where, $\alpha,\beta,\theta$ are three continuous parameters which describe the little group $W$.  $R\left( \theta  \right)$ is the two-dimensional rotation around $z$ axis;
\begin{displaymath}
R\left( \theta  \right) = \left( {\begin{array}{*{20}{c}}
{\cos \theta }&{\sin \theta }&0&0\\
{ - \sin \theta }&{\cos \theta }&0&0\\
0&0&1&0\\
0&0&0&1
\end{array}} \right)
\end{displaymath}
The matrix $S$ is a general Lorentz transformation, and it can be expressed as the following form:
\begin{equation} \label{eq:SS}
S\left( {\alpha ,\beta } \right) = \left( {\begin{array}{*{20}{c}}
u&v&{ - \alpha }&{q\alpha }\\
w&s&{ - \beta }&{q\beta }\\
x&y&{1 - \zeta }&{q\zeta }\\
z&r&{ - q\zeta }&{1 + {q^2}\zeta }
\end{array}} \right)
\end{equation}
Where, $q=p/E$ is a dimensionless real parameter; $q=1$ represents the massless standard momentum, and $q\neq1$ represents massive ones. Besides, we can obtain the following equation:
\[{\alpha ^2} + {\beta ^2} + \left( {1 - {q^2}} \right){\zeta ^2} - 2\zeta  = 0\]
The $\zeta$ satisfying this equation has two solutions, and only the following solution is taken:
\[\zeta  = \frac{1}{{1 - {q^2}}}\left[ {1 - \sqrt {1 - \left( {1 - {q^2}} \right)\left( {{\alpha ^2} + {\beta ^2}} \right)} } \right]\]

The former two columns of the matrix $S$ are still unknown, and it can be solved by the conditions of Lorentz transformation, which is  $g = {S^T}gS$ (Eq.2.3.5 in Ref.\cite{weinberg}), where $g$ is the Lorentz metric. In fact all the matrix elements can be solved through this equation. The following results are obtained:
\[\begin{array}{l}
s = \sqrt {1 - \left({1-{q^2}}\right){\beta ^2}} ;\quad v=\frac{{ - 1}}{s}\left( {1 - {q^2}}\right)\alpha\beta;\\
u = \sqrt {1 - \left( {1 - {q^2}} \right)\frac{{{\alpha ^2}}}{{{s^2}}}};\quad w = 0; \\
x = \frac{\alpha }{s}; \quad y = u\beta ;\quad z = xq;\quad r = yq
\end{array}\]

Note that: the obtained solutions are not unique, and here we only select one group among them. It can be verified that the different solutions corresponding to Lie algebra are equivalent to each other.

The infinitesimal transformation is conducted for the little group  $W\left( {\alpha ,\beta ,\theta } \right) = S\left( {\alpha ,\beta } \right)R\left( \theta  \right)$ to obtain its Lie algebra. That is to say, make $\alpha  \to 0,\beta  \to 0,\theta  \to 0$, simultaneously ignore second order and above items, and then obtain the following expression:
\[
W_{\ \nu }^\mu \left( {\theta ,\alpha ,\beta } \right) = I + \left( {\begin{array}{*{20}{c}}
0&\theta &{ - \alpha }&{q\alpha }\\
{ - \theta }&0&{ - \beta }&{q\beta }\\
\alpha &\beta &0&0\\
{q\alpha }&{q\beta }&0&0
\end{array}} \right)
\]
The above is the infinitesimal expression of the little group $W$ which keeps the standard momentum  $k = \left( {0,0,p,E} \right)$ invariant.

The infinitesimal unitary transformation in the Hilbert space which corresponds to the little group $W$ is as follows (p71 in Ref.\cite{weinberg}):
\[
U\left( {W\left( {\alpha ,\beta ,\theta } \right)} \right) = 1 + i\alpha A + i\beta B + i\theta {J_3}
\]
Where, $A = {J_2} + q{K_1}\quad B =  - {J_1} + q{K_2}$; $J_i$ are angular momentum operators; $K_i$ are boost operators. The commutation relations of the generators are as follows:
\begin{equation} \label{eq:angular}
\left[ {{J_3},A} \right] = iB;\ \left[ {B,{J_3}} \right] = iA;\ \left[ {A,B} \right] = i\left( {1 - {q^2}} \right){J_3}
\end{equation}

From the generators ($A, B, J_3$) of the unitary transformation $U(W)$ and their commutation relations (\ref{eq:angular}) (namely Lie algebra), it can be seen that: if $q=1$, this Lie algebra is the Lie algebra \emph{iso}(2) (eqs.2.5.35-37 in Ref.\cite{weinberg}) of the massless states; if $q=0$, it is the Lie algebra \emph{o}(3) (angular momentums) of the massive states. That is to say, as the particle mass tends to 0, namely, $q$ tends to 1, the Lie algebra of the little group $W$ contracts from \emph{o}(3) to \emph{iso}(2). As the mass becomes 0, the Lie algebra has essential changes.

\section{Kinematic proof of massless particles having only two helicity states}

When $q\neq1$, we make
\begin{equation} \label{eq:angu_trans}
{J'_1} = \frac{A}{{\sqrt {1 - {q^2}} }},\quad {J'_2} = \frac{B}{{\sqrt {1 - {q^2}} }},\quad {J'_3} = {J_3}
\end{equation}
And then the above commutation relations (\ref{eq:angular}) become
\[\left[ {{J'_3},{J'_1}} \right] = i{J'_2};\quad \left[ {{J'_2},{J'_3}} \right] = i{J'_1};\quad \left[ {{J'_1},{J'_2}} \right] = i{J'_3}\]
These are the commutation relations of the angular momentums, and thus the Lie algebra (\ref{eq:angular}) is \emph{o}(3). Obviously it is correct, because when $q\neq1$, the standard momentum $k = \left( {0,0,p,E} \right)$  is the situation of massive particles, and its little group is naturally SO(3).

It needs to note that: when the particle mass tends to 0, the denominators in Eq.(\ref{eq:angu_trans}) are singular; when the eigenvalues of the operators $A$ and $B$ are not 0, it makes the eigenvalues of $J'_{1,2}$ tend to infinity. Generally infinity values are unobservable in physics, so the one-particle states which make the eigenvalues of $J'_{1,2}$ tend to infinity are nonphysical. Therefore, only the states which make the eigenvalues of the operators $A$ and $B$ be 0 are physical, and then these states make the eigenvalues of $J'_{1,2}$ be also 0. It needs to note that, here we do \emph{not} need any experimental hypothesis (this is different from p71-72 in Ref.\cite{weinberg}) and just admit: the states which make the eigenvalues of $J'_{1,2}$ tend to infinity are nonphysical, thereby obtaining that physical states are the one-particle states which make the eigenvalues of the operators $A$ and $B$ be 0. The obtained conclusions are the same as those of Eq.2.5.38 in Ref.\cite{weinberg}. However, any experimental hypothesis is not needed.

Below we discuss the group representations. The representation \cite{wigner} of the SO(3) group which corresponds to Lie algebra (\ref{eq:angular}) is as follows:

\end{multicols}
\begin{equation} \label{eq:rep_mass}
\begin{array}{l}
D_{\sigma '\sigma }^j\left( W \right) = D_{\sigma '\sigma }^j\left( {\alpha \beta \gamma } \right) = \left\langle {{\Psi _{p\sigma '}}} \right|\exp \left( { - i\alpha {J_3}} \right)\exp \left( { - i\beta B} \right)\exp \left( { - i\gamma {J_3}} \right)\left| {{\Psi _{p\sigma }}} \right\rangle \\
\quad \quad  = \left\langle {{\Psi _{p\sigma '}}} \right|\exp \left( { - i\alpha {{J'}_3}} \right)\exp \left( { - i\beta \sqrt {1 - {q^2}} {{J'}_2}} \right)\exp \left( { - i\gamma {{J'}_3}} \right)\left| {{\Psi _{p\sigma }}} \right\rangle \\
\quad \quad  = \sum\limits_n^{} {{{\left( { - 1} \right)}^n}\dfrac{{\sqrt {\left( {j - \sigma } \right)!\left( {j + \sigma } \right)!\left( {j - \sigma '} \right)!\left( {j + \sigma '} \right)!} }}{{\left( {j + \sigma ' - n} \right)!\left( {j - \sigma  - n} \right)!n!\left( {n + \sigma  - \sigma '} \right)!}}\exp \left( { - i\sigma '\alpha } \right) \times } \\
\quad \quad \;\;{\left( {\cos \frac{{\beta \sqrt {1 - {q^2}} }}{2}} \right)^{2j + \sigma ' - \sigma  - 2n}}{\left( {\sin \frac{{\beta \sqrt {1 - {q^2}} }}{2}} \right)^{2n - \sigma ' + \sigma }}\exp \left( { - i\sigma \gamma } \right)
\end{array}
\end{equation}
\begin{multicols}{2}
Where, $j$ is the spin quantum number; the value range of indexes of the matrix is $\sigma ,\sigma ' = j,j - 1, \cdots , - j + 1, - j$ ; the value range of the summation index integer $n$ is $n \ge 0,\quad n \ge \sigma ' - \sigma ,\quad n \le j - \sigma ,\quad n \le j + \sigma '$ ; parameters $\alpha,\beta$ and $\gamma$ are Eulerian angles. When $q=0$, the representation (\ref{eq:rep_mass}) is just the representation of the SO(3) group in Eq.(\ref{eq:massive1}).

As the particle mass tends to 0, the physical one-particle states are only considered which make the eigenvalues of the operators $A$ and $B$ be 0, and the above representation (\ref{eq:rep_mass}) of SO(3) group contracts to the following finite-dimensional representation:
\end{multicols}
\begin{equation} \label{eq:rep_nomass}
\begin{array}{l}
D_{\sigma '\sigma }\left( W \right) = \left\langle {{\Psi _{p\sigma '}}} \right|\exp \left( { - i\alpha {J_3}} \right)\exp \left( { - i\beta B} \right)\exp \left( { - i\gamma {J_3}} \right)\left| {{\Psi _{p\sigma }}} \right\rangle \\
\quad \quad  = \exp \left( { - i\sigma '\alpha } \right)\left\langle {{\Psi _{p\sigma '}}} \right|\left( {1 + \sum\limits_{r = 1}^{\infty} {\frac{{{{\left( { - i\beta B} \right)}^r}}}{{r!}}} } \right)\left| {{\Psi _{p\sigma }}} \right\rangle \exp \left( { - i\sigma \gamma } \right)\\
\quad \quad  = {\delta _{\sigma '\sigma }}\exp \left( { - i\sigma \left( {\alpha  + \gamma } \right)} \right)
\end{array}
\end{equation}
\begin{multicols}{2}
Eq.(\ref{eq:rep_nomass}) is just the representation of the SO(2) group, namely the representation of the group in Eq.(\ref{eq:massless1}). This proves that: in the physical level, as the particle's mass tends to 0, representation (\ref{eq:rep_mass}) contracts to representation (\ref{eq:rep_nomass}), and thus Eq.(\ref{eq:massive1}) is continuously changed into Eq.(\ref{eq:massless1}); the $2j+1$ massive one-particle states are degenerated to two states of the massless particles.

It is needed to prove that the massive one-particle states converge to massless states when the particle's mass tends to zero. First, we discuss the changes of the vector particles whose spin is 1 when the mass tends to 0. According to the helicity of the vector particles, all the states  of the vector particles can be divided into the following three parts:${\Psi _{p1}},{\Psi _{p0}},{\Psi _{p - 1}}$. The set of the states which make the eigenvalues of the operators $A$ and $B$ be 0 is some subset of all the above states, and such subset has two kinds: \textcircled{1} ${\Psi _{p1}},{\Psi _{p - 1}}$; \textcircled{2} ${\Psi _{p0}}$. Because the matrix elements of ladder operators  ${J'_ \pm } = {J'_1} \pm i{J'_2}$ (which includes the operators $A$ and $B$) are all 0 in the above mentioned any subset, those eigenvalues are naturally also 0, and these two subsets may be physical states. When the particle mass tends to 0, which subset of the above is physical? If the subset \textcircled{2} is physical, the helicity of all the observed particles will be 0, and these particles should be considered as scalar particles while not vector particles. Therefore, the physical states can only be the subset \textcircled{1}, and the helicity observed in this subset are $\pm1$. In this way, it can clearly see which states are physical and which states are nonphysical as the particle mass tends to 0.

For the massless particles whose spin is other values, the same discussion can be conducted. Below we conducted explanations with particles spin 2. The subsets of the states which make the eigenvalues of the operators $A$ and $B$ be 0 have the following four kinds: \textcircled{1}${\Psi _{p2}},{\Psi _{p - 2}}$; \textcircled{2}${\Psi _{p1}},{\Psi _{p - 1}}$; \textcircled{3}${\Psi _{p0}}$; \textcircled{4}${\Psi _{p2}},{\Psi _{p - 2}},{\Psi _{p0}}$. The helicity shown by the subsets\textcircled{2} and \textcircled{3} is not 2, and thus they are excluded. The subset \textcircled{4} contains two invariant subspaces, namely \textcircled{1} and \textcircled{3}, and thus the Lorentz group representation carried by the subset \textcircled{4} is reducible. Generally this is impossible in physics. Therefore, only the subset \textcircled{1} is physically allowed, and the helicity has only two states. In this way, as the mass tends to 0, the states with mass are degenerated into the massless states.

Below, we prove it again from dynamic perspectives. Let us consider the particle's momentum is in the 3-direction. The polarization vectors $\epsilon ^{\mu} \left({p,\lambda}\right) $ (p154-158 in Ref.\cite{greiner})  for the massive vector field are (where $p$ is the four momentum)
\[\begin{array}{l}
{p^\mu } = \left( {0,0,\sqrt {{E^2} - {m^2}} ,E} \right)\\
{\varepsilon ^\mu }\left( {p,0} \right) = \dfrac{{{p^\mu }}}{m}\\
{\varepsilon ^\mu }\left( {p,1} \right) = \left( {1,0,0,0} \right)\\
{\varepsilon ^\mu }\left( {p,2} \right) = \left( {0,1,0,0} \right)\\
{\varepsilon ^\mu }\left( {p,3} \right) = \dfrac{1}{m}\left( {0,0,E,\sqrt {{E^2} - {m^2}} } \right)
\end{array}\]
Correspondingly, the polarization vectors for the massless vector field are
\[\begin{array}{l}
{p^\mu } = \left( {0,0,E,E} \right)\\
{\varepsilon ^\mu }\left( {p,0} \right) = \left( {0,0,0,1} \right)\\
{\varepsilon ^\mu }\left( {p,1} \right) = \left( {1,0,0,0} \right)\\
{\varepsilon ^\mu }\left( {p,2} \right) = \left( {0,1,0,0} \right)\\
{\varepsilon ^\mu }\left( {p,3} \right) = \left( {0,0,1,0} \right)
\end{array}\]

The transverse polarization vectors of massive and massless particles are the same, and they are no relation with particles' masses. The creation operators (p162 in Ref.\cite{greiner}) of massive and massless particles are
\[a_{p\lambda }^\dag  = i\int {{d^3}x\frac{{{e^{ - ip \cdot x}}}}{{\sqrt {2{p^0}{{\left( {2\pi } \right)}^3}} }}{\varepsilon ^\mu }{{\left( {p,\lambda } \right)}^*}\left( {{\partial _0}{A_\mu }\left( x \right) + i{p^0}{A_\mu }\left( x \right)} \right)} \]
Because the transverse polarization vectors ($\epsilon ^{\mu} \left({p,\lambda}\right), \lambda=1,2$) of massive and massless particles are the same, the transverse creation operators of massive and massless particles are the same when the particles mass tend to zero.

The transverse one-particle states ($\left| 1 \right\rangle  = \int {d\tilde pf\left( p \right)a_{p\lambda }^\dag } \left| 0 \right\rangle, \quad \lambda  = 1,2$) of massive particles converge to those of massless ones when particles masses tend to zero.

Therefore, one-particle states generated by creation operators are the same when $q \to 1$. The proof is from dynamic perspectives. We do not discuss the scalar and longitudinal one-particle states generated by creation operators which are nonphysical for massless particles.

Now we talk something about Pauli-Lubanski vector (${W^\mu } = -\frac{1}{2}{\varepsilon ^{\mu \nu \rho \sigma }}{P_\nu }{J_{\rho \sigma }}$). For massless particles, the square of Pauli-Lubanski vector (a Casimir operator of Poincar\'{e} group and ISO(2) group) does not equal zero in mathematics\cite{Maggiore}.
\[{W^\mu }{W_\mu } = {E^2}\left[ {{{\left( {{J_2} + {K_1}} \right)}^2} + {{\left( { - {J_1} + {K_2}} \right)}^2}} \right] = {E^2}\left( {{A^2} + {B^2}} \right)\]
For physical states which make eigenvalues of operators $A$ and $B$ be zero, $W^2=0$. Here, we do not need the explanation like Ref.\cite{weinberg,Maggiore}.

\section{Conclusions}


As the particle mass tends to 0, when the eigenvalues of the operators $A$ and $B$ are not 0, the eigenvalues of $J'_{1,2}$ tend to infinity, and this is nonphysical. Therefore only those states which make the eigenvalues of the operators $A$ and $B$ be 0 are physical. Such states make representation (\ref{eq:rep_mass}) degenerate to representation (\ref{eq:rep_nomass}), and thus Eq.(\ref{eq:massive1}) is continuously changed to Eq.(\ref{eq:massless1}). In this way, any experimental hypotheses are not needed, and from the perspectives of pure theory and kinematics it proves that: when the particle mass tends to 0, the $2j+1(j>0)$ helicity states of massive particles degenerate to 2 helicity states of the massless particles.

\end{multicols}

\vspace{-1mm}
\centerline{\rule{80mm}{0.1pt}}
\vspace{2mm}

\begin{appendix}
\begin{center}
    {\bf APPENDIX}
\end{center}

\section{Contraction from SO(n) to ISO(n-1), and SO(n,m) to ISO(n,m-1)}

In Ref.\cite{gilmore}, the In\"{o}n\"{u}-Wigner contraction is applied to $SO(n)\to ISO(n-1)$ and $SO(3,2)\to ISO(3,1)$. The contraction method \cite{czachor,kim} is easily generalized to $SO(n)\to ISO(n-1)$ and $SO(3,2)\to ISO(3,1)$.

First, we talk $SO(3)\to ISO(2)$ again. The little group $W$ (which satisfies $W_{\ \nu }^\mu {k^\nu } = {k^\mu }$) (where $k = \left( {0,0,q,1} \right),\quad 0 \le q \le 1$) can be split into two parts $W=SR$ where $R$ is rotation group $SO(2)$. How to comprehend the action of the matrix $S$ is the key point. The zero components in standard momentum $k$ are transformed by the matrix $R$, and non-zero components in $k$ are transformed by the matrix $S$. The action of last two elements of the first row of matrix $S$ (Eq.(\ref{eq:SS})) is that: they transform the third component $q$ in $k$ to the first place in $k$ (on the left side of "$=$"), and then transform the fourth part $1$ and cancel them. The decomposition is as follows. In the appendix, we only consider infinitesimal transformations.
\[
\begin{array}{l}
{k^\mu } = {S_{ \nu }^\mu}{k^\nu } = I\left( {\begin{array}{*{20}{c}}
0\\
0\\
q\\
1
\end{array}} \right) + \left( {\begin{array}{*{20}{c}}
0&0&{ - {\alpha _1}}&{q{\alpha _1}}\\
0&0&{ - {\alpha _2}}&{q{\alpha _2}}\\
{{\alpha _1}}&{{\alpha _2}}&0&0\\
{q{\alpha _1}}&{q{\alpha _2}}&0&0
\end{array}} \right)\left( {\begin{array}{*{20}{c}}
0\\
0\\
q\\
1
\end{array}} \right)\\
\quad  = \left( {\begin{array}{*{20}{c}}
0\\
0\\
q\\
1
\end{array}} \right) + \left[ {{\alpha _1}\left( {\begin{array}{*{20}{c}}
0&0&{ - 1}&q\\
0&0&0&0\\
1&0&0&0\\
q&0&0&0
\end{array}} \right) + {\alpha _2}\left( {\begin{array}{*{20}{c}}
0&0&0&0\\
0&0&{ - 1}&q\\
0&1&0&0\\
0&q&0&0
\end{array}} \right)} \right]\left( {\begin{array}{*{20}{c}}
0\\
0\\
q\\
1
\end{array}} \right)\\
\quad  = \left( {\begin{array}{*{20}{c}}
0\\
0\\
q\\
1
\end{array}} \right) + \left\{ \begin{array}{l}
\left[ {{\alpha _1}\left( {\begin{array}{*{20}{c}}
0&0&{ - 1}&0\\
0&0&0&0\\
1&0&0&0\\
0&0&0&0
\end{array}} \right) + {\alpha _1}\left( {\begin{array}{*{20}{c}}
0&0&0&q\\
0&0&0&0\\
0&0&0&0\\
q&0&0&0
\end{array}} \right)} \right] + ... \\
\end{array} \right\}\left( {\begin{array}{*{20}{c}}
0\\
0\\
q\\
1
\end{array}} \right)\\
\quad  = \left( {\begin{array}{*{20}{c}}
0\\
0\\
q\\
1
\end{array}} \right) + \left[ {{\alpha _1}\left( {\begin{array}{*{20}{c}}
{ - q}\\
0\\
0\\
0
\end{array}} \right) + {\alpha _1}\left( {\begin{array}{*{20}{c}}
q\\
0\\
0\\
0
\end{array}} \right) + {\alpha _2}\left( {\begin{array}{*{20}{c}}
0\\
{ - q}\\
0\\
0
\end{array}} \right) + {\alpha _2}\left( {\begin{array}{*{20}{c}}
0\\
q\\
0\\
0
\end{array}} \right)} \right] = \left( {\begin{array}{*{20}{c}}
0\\
0\\
q\\
1
\end{array}} \right)
\end{array}
\]

Now we know the action of matrix $S$, and we can directly write down it without solving equations such as in section 2.
The situation $n=4$ is discussed here. We set standard momentum in five-dimensional space $k = \left( {0,0,0,q,1} \right),\quad 0 \le q \le 1$. When $q=0$, the little group $W$ is SO(4).  $W$ can be split into the product of two transformations, namely  $W\left( {{\alpha _i}{\beta _j}} \right) = S\left( {{\beta _j}} \right)R\left( {{\alpha _i}} \right),\quad i,j = 1,2,3$, where $R\left( {{\alpha _i}} \right)$ is SO(3) group. The little group's Lie algebra can easily be written down

\[{W_{\ \nu }^\mu } = I + \left( {\begin{array}{*{20}{c}}
0&{ - {\alpha _3}}&{{\alpha _2}}&{ - {\beta _1}}&{q{\beta _1}}\\
{{\alpha _3}}&0&{ - {\alpha _1}}&{ - {\beta _2}}&{q{\beta _2}}\\
{ - {\alpha _2}}&{{\alpha _1}}&0&{ - {\beta _3}}&{q{\beta _3}}\\
{{\beta _1}}&{{\beta _2}}&{{\beta _3}}&0&0\\
{q{\beta _1}}&{q{\beta _2}}&{q{\beta _3}}&0&0
\end{array}} \right) = I + \sum\limits_{i = 1}^3 {\left( {{\alpha _i}{A_i} + {\beta _i}{B_i}} \right)} \]

The six generators are

\[\begin{array}{l}
{A_1} = \left( {\begin{array}{*{20}{c}}
0&0&0&0&0\\
0&0&{ - 1}&0&0\\
0&1&0&0&0\\
0&0&0&0&0\\
0&0&0&0&0
\end{array}} \right)\quad
{A_2} = \left( {\begin{array}{*{20}{c}}
0&0&1&0&0\\
0&0&0&0&0\\
{ - 1}&0&0&0&0\\
0&0&0&0&0\\
0&0&0&0&0
\end{array}} \right)\quad
{A_3} = \left( {\begin{array}{*{20}{c}}
0&{ - 1}&0&0&0\\
1&0&0&0&0\\
0&0&0&0&0\\
0&0&0&0&0\\
0&0&0&0&0
\end{array}} \right)\\
{B_1} = \left( {\begin{array}{*{20}{c}}
0&0&0&{ - 1}&q\\
0&0&0&0&0\\
0&0&0&0&0\\
1&0&0&0&0\\
q&0&0&0&0
\end{array}} \right) \quad
{B_2} = \left( {\begin{array}{*{20}{c}}
0&0&0&0&0\\
0&0&0&{ - 1}&q\\
0&0&0&0&0\\
0&1&0&0&0\\
0&q&0&0&0
\end{array}} \right)\quad
{B_3} = \left( {\begin{array}{*{20}{c}}
0&0&0&0&0\\
0&0&0&0&0\\
0&0&0&{ - 1}&q\\
0&0&1&0&0\\
0&0&q&0&0
\end{array}} \right)
\end{array}
\]

The commutation relations are
\[
\left[ {{A_i},{A_j}} \right] = {\varepsilon _{ijk}}{A_k}\quad
\left[ {{A_i},{B_j}} \right] = {\varepsilon _{ijk}}{B_k}\quad
\left[ {{B_i},{B_j}} \right] = \left( {1 - {q^2}} \right){\varepsilon _{ijk}}{A_k}
\]

When $q=0$, the commutation relations are SO(4)'s ones. When $q\to1$, the relations become ISO(3)'s (p209 in Ref.\cite{gilmore}).

In the following, we talk about $SO(m,n)$ group. It is not difficult to generalize the method to $SO(3,2)\to ISO(3,1)$ (generators are in p86 of Ref.\cite{gilmore}) and so on. We set the standard momentum $k = \left( {0,0,0,0,q,1} \right),\quad 0 \le q \le 1$. The Lie algebra of little group $W$ is
\[W = \left( {\begin{array}{*{20}{c}}
1&{{\alpha _3}}&{ - {\alpha _2}}&{{\beta _1}}&{{\gamma _1}}&{ - q{\gamma _1}}\\
{ - {\alpha _3}}&1&{{\alpha _1}}&{{\beta _2}}&{{\gamma _2}}&{ - q{\gamma _2}}\\
{{\alpha _2}}&{ - {\alpha _1}}&1&{{\beta _3}}&{{\gamma _3}}&{ - q{\gamma _3}}\\
{{\beta _1}}&{{\beta _2}}&{{\beta _3}}&1&{{\gamma _4}}&{ - q{\gamma _4}}\\
{{\gamma _1}}&{{\gamma _2}}&{{\gamma _3}}&{ - {\gamma _4}}&1&0\\
{q{\gamma _1}}&{q{\gamma _2}}&{q{\gamma _3}}&{ - q{\gamma _4}}&0&1
\end{array}} \right)\]

The generators are
\[
\begin{array}{l}
{J_1} = \left( {\begin{array}{*{20}{c}}
0&0&0&0&0&0\\
0&0&1&0&0&0\\
0&{ - 1}&0&0&0&0\\
0&0&0&0&0&0\\
0&0&0&0&0&0\\
0&0&0&0&0&0
\end{array}} \right);{J_2} = \left( {\begin{array}{*{20}{c}}
0&0&{ - 1}&0&0&0\\
0&0&0&0&0&0\\
1&0&0&0&0&0\\
0&0&0&0&0&0\\
0&0&0&0&0&0\\
0&0&0&0&0&0
\end{array}} \right);{J_3} = \left( {\begin{array}{*{20}{c}}
0&1&0&0&0&0\\
{ - 1}&0&0&0&0&0\\
0&0&0&0&0&0\\
0&0&0&0&0&0\\
0&0&0&0&0&0\\
0&0&0&0&0&0
\end{array}} \right)\\
{K_1} = \left( {\begin{array}{*{20}{c}}
0&0&0&1&0&0\\
0&0&0&0&0&0\\
0&0&0&0&0&0\\
1&0&0&0&0&0\\
0&0&0&0&0&0\\
0&0&0&0&0&0
\end{array}} \right);{K_2} = \left( {\begin{array}{*{20}{c}}
0&0&0&0&0&0\\
0&0&0&1&0&0\\
0&0&0&0&0&0\\
0&1&0&0&0&0\\
0&0&0&0&0&0\\
0&0&0&0&0&0
\end{array}} \right);{K_3} = \left( {\begin{array}{*{20}{c}}
0&0&0&0&0&0\\
0&0&0&0&0&0\\
0&0&0&1&0&0\\
0&0&1&0&0&0\\
0&0&0&0&0&0\\
0&0&0&0&0&0
\end{array}} \right)\\
{P_1} = \left( {\begin{array}{*{20}{c}}
0&0&0&0&1&{ - q}\\
0&0&0&0&0&0\\
0&0&0&0&0&0\\
0&0&0&0&0&0\\
1&0&0&0&0&0\\
q&0&0&0&0&0
\end{array}} \right)\quad {P_2} = \left( {\begin{array}{*{20}{c}}
0&0&0&0&0&0\\
0&0&0&0&1&{ - q}\\
0&0&0&0&0&0\\
0&0&0&0&0&0\\
0&1&0&0&0&0\\
0&q&0&0&0&0
\end{array}} \right)\\
{P_3} = \left( {\begin{array}{*{20}{c}}
0&0&0&0&0&0\\
0&0&0&0&0&0\\
0&0&0&0&1&{ - q}\\
0&0&0&0&0&0\\
0&0&1&0&0&0\\
0&0&q&0&0&0
\end{array}} \right)\quad {P_4} = H = \left( {\begin{array}{*{20}{c}}
0&0&0&0&0&0\\
0&0&0&0&0&0\\
0&0&0&0&0&0\\
0&0&0&0&1&{ - q}\\
0&0&0&{ - 1}&0&0\\
0&0&0&{ - q}&0&0
\end{array}} \right)
\end{array}
\]
The communication relations are
\[\begin{array}{*{20}{l}}
{\left[ {{J_i},{J_j}} \right] =  - {\varepsilon _{ijk}}{J_k}}&{\left[ {{J_i},{K_j}} \right] =  - {\varepsilon _{ijk}}{K_k}}&{\left[ {{K_i},{K_j}} \right] = {\varepsilon _{ijk}}{J_k}}&{\left[ {{J_i},{P_j}} \right] =  - {\varepsilon _{ijk}}{P_k}}\\
{\left[ {H,{K_i}} \right] = -{P_i}}&{\left[ {{P_i},{K_j}} \right] = - H{\delta _{ij}}}&{\left[ {{P_i},{P_j}} \right] = \left( {1 - {q^2}} \right){\varepsilon _{ijk}}{J_k}}&{\left[ {H,{P_i}} \right] = \left( {1 - {q^2}} \right){K_i}}
\end{array}\]
When $q=0$, the commutation relations are SO(3,2)'s ones. When $q\to1$, the relations become ISO(3,1)'s ones (namely Poincar\'{e} group).

\end{appendix}

\vspace{2mm}
\centerline{\rule{80mm}{0.1pt}}
\vspace{2mm}
\begin{multicols}{2}

\end{multicols}

\clearpage

\end{document}